\documentclass[10pt,twocolumn,letter]{IEEEtran}
\usepackage{times,bm}
\usepackage{bbold}
\usepackage{bbm}
\usepackage{url}
\usepackage{indentfirst}
\usepackage{cite}
\usepackage{subfigure}
\usepackage{amsmath}
\usepackage{multicol}
\usepackage{multirow}
\usepackage{amsfonts}
\usepackage{ textcomp, comment}
\usepackage{upgreek}
\usepackage{times}
\usepackage[dvips]{graphicx}
\usepackage[usenames]{color}
\usepackage{epsf}
\usepackage{psfrag}

\usepackage{flushend}

\newcommand{\ben}{\begin{enumerate}}
\newcommand{\een}{\end{enumerate}}
\newcommand{\be}{\begin{equation}}
\newcommand{\ee}{\end{equation}}
\newcommand{\bea}{\begin{eqnarray}}
\newcommand{\eea}{\end{eqnarray}}
\newcommand{\bc}{\begin{cases}}
\newcommand{\ec}{\end{cases}}
\newcommand{\bi}{\begin{itemize}}
\newcommand{\ei}{\end{itemize}}

\newcommand{\bre}{\begin{bf} \begin{color}{BrickRed} }
\newcommand{\ere}{\end{color} \end{bf}}
\DeclareMathOperator*{\argmax}{\arg\!\max}
\usepackage{epsf}
\usepackage{psfrag}

\graphicspath{%
	{./}
}
\begin{document}
\sloppy

\title{Estimating the number of receiving nodes in \\802.11 networks via machine learning techniques}

\author{\authorblockN{{\bf Davide {Del~Testa}}\IEEEauthorrefmark{1}, {\bf Matteo Danieletto}\IEEEauthorrefmark{3}, {\bf Giorgio Maria Di Nunzio}\IEEEauthorrefmark{1}, and {\bf Michele Zorzi}\IEEEauthorrefmark{1}}\\
\IEEEauthorblockA{\IEEEauthorrefmark{1}Department of Information Engineering (DEI), University of Padova, Italy \\ \{deltesta, dinunzio, zorzi\}@dei.unipd.it}\\
\IEEEauthorblockA{\IEEEauthorrefmark{3}Qualcomm Institute, University of California San Diego -- La Jolla, CA, USA \\ mdanieletto@eng.ucsd.edu}
}

\maketitle

%\vspace*{-2truecm}

\thispagestyle{empty}

\begin{abstract}
Nowadays, most mobile devices are equipped with multiple wireless interfaces, causing an emerging research interest in device to device (D2D) communication: the idea behind the D2D paradigm is to exploit the proper interface to directly communicate with another user, without traversing any network infrastructure. 
A first issue related to this paradigm consists in the need for a coordinator, called controller, able to decide when activating a D2D connection is appropriate and eventually able to manage such connection. In this view, the paradigm of Software Defined Networking (SDN), can be exploited both to handle the data flows among the devices and to interact directly with every device. 
 This work is focused on a scenario where a device is selected by the SDN controller, in order to become the master node of a WiFi-Direct network. The remaining nodes, called clients, can exchange data with other nodes through the master. The objective is to infer, through different Machine Learning approaches, the number of nodes actively involved in receiving data, exploiting only the information available at the client side and without modifying any standard communication protocol. The information about the number of client nodes is crucial when, \emph{e.g.}, a user desires a precise prediction of the transmission estimated time of arrival (ETA) while downloading a file.
\end{abstract}

\section{Introduction}
\label{sec:introduction}

Nowadays, more than two billions of smartphones are active in the global market \cite{smartphone}. One of the most important features of a smartphone is to provide an Internet connection, allowing each user to access/share their data, \emph{e.g.}, e-mails, photos, videos, \emph{etc}.

Two main factors usually affect a smartphone's capability to stay connected and exchange data: the battery life and the amount of data provided by the cellular carrier subscription. These can be addressed by exploiting heterogeneous wireless interface cards (WNICs) such as WiFi and Bluetooth, in order to simultaneously minimize the battery consumption and save a portion of the data transferred by the cellular network~\cite{d2dbattery}. However, this capability to automatically exploit heterogeneous WNICs is not yet supported by any smartphone service; indeed, if a user wants to connect her/his device to a WiFi connection, she/he has to manually set it up, provided that she/he either knows the SSID and the WiFi encryption parameters, or decides whether an open WiFi connection is worth trusting.

In the last few years, researchers in the Software Defined Networking (SDN) area have been focused on creating a software protocol suite that is decoupled from the underlying 
hardware. This entails the possibility to, \emph{e.g.}, prototype new protocols~\cite{mckeown2008openflow} and create network virtualization and traffic isolation (the reader can find a complete overview in \cite{SDN_SURVEY}). 
The SDN paradigm goes in this direction by decoupling the network traffic into two planes: the control plane and the data plane. The control plane aims at monitoring routers, switches and Access Points (APs) to handle the network behavior and report any kind of useful information to a centralized unit, called controller. The controller can establish one or more data traffic flows among the network hosts by using the information gathered from the network status. These can be configured through some predefined Quality of Service network policies, such as offloading, number of hops between two users or energy consumption for the User Equipment (UE). Therefore, a controller can exchange information and manipulate the switch/router behavior by employing an ad hoc protocol like OpenSwitch~\cite{open_switch}. On the other hand, the data plane handles all the data traffic generated by the host, which is completely separated from the control plane.

A further step in this direction can be undertaken when the SDN concept is paired with wireless technologies, as explained in \cite{openroads, sdn_wifi_dense_2}. For the above-mentioned purpose, a number of real WiFi SDN testbed implementations have been presented, \emph{e.g.},  OpenRoads~\cite{openroads}, Odin~\cite{odin}, OFRewind~\cite{ofrewind} and CARMEN~\cite{carmen}. Among these testbeds, OpenRoads and CARMEN are the most promising: the former was the first testbed to exploit OpenSwitch and to give the possibility to use a WiFiMax connection to test traffic offloading between two heterogeneous wireless interfaces; the latter has $50$ WiFi nodes in fixed positions deployed in a university campus area.
%where each node can work either as an access point device or in ad hoc mode, exploiting a wired connection in order to create the control plane and share network status information with the controller. 
Moreover, three features characterize CARMEN from the other testbeds: it is able to collect all network parameters from datalink layer to application layer, it can exploit 
Android smartphone devices for mobility analysis, and it provides each node with at least two IEEE 802.11 WNICs.
% a IEEE 802.15.4 WNIC and least two IEEE 802.11 WNICs.

Finally, a further possibility is to integrate the device to device (D2D) communication into the SDN concept. D2D communication arises when two mobile nodes are able to communicate directly without traversing any infrastructure, \emph{i.e.}, a base station (more details regarding the D2D concept can be found in \cite{asadi2014survey}). D2D can generally exploit cellular spectrum (\emph{i.e.}, inband) or unlicensed spectrum (\emph{i.e.}, outband). In a D2D architecture the device network interfaces are expected to be overseen/controlled by a central entity \emph{e.g.}, the SDN controller.
%; second, the D2D user is typically only allowed to act autonomously only when the cellular infrastructure is unavailable.
Outband allows to eliminate the interference issues between D2D and cellular links, at the cost of an extra interface to be used, as it usually adopts other wireless technologies, such as WiFi, Bluetooth or IEEE 802.15.4. 
Therefore, the SDN paradigm applied to D2D makes it possible to dynamically switch between different interfaces during transmission, based on predefined policies, \emph{e.g.}, data offloading, traffic balance and battery energy consumption. Hence, it is possible to design and implement a heterogeneous network environment where the device can exploit both network interfaces through the D2D concept in an unlicensed band. The idea to couple SDN and D2D has already been studied in \cite{cellular_offloading,sdn_heterogenus1,sdn_heterogenus2}, as well as the exploitation of the WiFi-Direct paradigm to set up a D2D network~\cite{wifi_direct1,wifi_direct2} with several devices. 

In this work, we exploit CARMEN to generate an SDN scenario, where a D2D network is configured with WiFi Direct. The SDN controller, installed into the centralized radio access network (C-RAN), exploits an LTE connection to create a D2D network and to select its master node. The aim is to predict client-side information related to the nodes generating data traffic in the same D2D network. This procedure should: i) not interact with any SDN controller; ii) be completely transparent for the user; and iii) avoid to alter any network or SDN protocol (\emph{e.g.}, OpenSwitch/OpenFlow). 
%In our previous work~\cite{deltesta_globecomm}, an SDN WiFi scenario was set up where a node was designated as access point (\emph{i.e.}, master in case of WiFi Direct scenario) and every others nodes (acting as a client) collected several network parameters in order to predict an accurate Estimated Time of Arrival (ETA). 
In our previous work~\cite{deltesta_globecomm}, we considered an SDN WiFi scenario with a node designated as the access point (\emph{i.e.}, master in case of a WiFi Direct scenario) and a varying number of nodes acting as clients. These nodes collected several network parameters in order to accurately predict an Estimated Time of Arrival (ETA). There, we used the number of simultaneously receiving nodes as an input datum, assuming that this information was available at each device. However, this assumption is unrealistic in practice, as the number of receiving nodes is only known to the transmitter (access point) which, in general, does not share this value unless a modification of the transmission protocol is introduced. Moreover, two reasons prevent the SDN controller from acknowledging how many nodes are actively involved in data reception inside the WiFi D2D network.
First, the master is acting as D2D gateway, and the LTE C-RAN side is completely blind regarding who is generating the traffic data. Second, the slave nodes use their local OpenFlow table already configured for D2D, and do not report any information to the SDN controller through the LTE connection.
%by means of a number of machine learning techniques, including Support Vector Regression (SVR) machines and Restricted Boltzmann Machines (RBMs). 

In this paper, we show how the number of the active UEs can be estimated accurately by the receiving nodes, by means of Machine Learning (ML) techniques and by taking as input only those network parameters available at the client side, in order to respect the requirement of not modifying any protocol. In order to accomplish this, we measure the amount of time needed for the transmission of the first fraction of a file from the AP to a receiving node, together with other information, \emph{e.g.}, the distance of each node to the AP and its transmission power.  
%As reported in Section~\ref{sec:results}, mistakes in nodes estimation cause a high ETA prediction error, hence, in this paper, we compare various techniques giving the highest classification performance on the number of active nodes.
%

The rest of the paper is organized as follows. In Section~\ref{sec:testbeddataset} we describe the testbed and the dataset employed in our experiments.
In Section~\ref{sec:ml} we offer an overview on the ML techniques considered to estimate the number of active nodes. Section~\ref{sec:results} presents the experimental evaluation of the ML techniques used to predict the number of active nodes. Finally, Section~\ref{sec:conclusions} concludes the paper and proposes some future work.

\section{Testbed and dataset}\label{sec:testbeddataset}
In this section, we describe how the data was collected and characterize the dataset in terms of dimensionality and network parameters.

\subsection{WiFi Testbed}
We conducted a thorough experimental data gathering to measure different network realizations and their related outputs. The data was collected using CARMEN \cite{carmen}, composed of 50 Alix miniPCs model 3d2 with WiFi driver \textit{$ath9k$} and five Nexus 7 devices with the \textit{$ath9k\_htc$} WiFi driver. In order to minimize the setup time of a specific WiFi network and quickly collect its performance, this testbed was associated with a software able to automatically gather data and change the network configurations. 
This software can collect the TCP/IP stack parameters, starting from the MAC sublayer up to the Application layer, and to remotely set the WiFi transmission channel and power, as well as the distance among the Alix nodes and the transmitter (more details can be found in \cite{carmen}). In addition, we exploited the GNU program \textit{cron} to schedule all experiments at different times of the day, in order to automate the entire data collection phase.
Being the data transmissions dependent on the channel status, several WiFi network setups were considered, by changing all possible parameters in order to modify the scenario and consequently the ETA measurements. Since the testbed is located inside our department, other WiFi nodes transmitting data can occupy the WiFi transmission channel, generating interference during the experiments. Our idea to mitigate this issue was to measure transmission data multiple times over different WiFi frequencies (channels) and times of the day (\emph{i.e.}, morning, afternoon and night). 

\subsection{Dataset characterization}
The final dataset has been created by varying, for each time of the day, the following network and topology parameters: transmission power \{$0$ dBm, $5$ dBm, $10$ dBm, $15$ dBm, $20$ dBm\} , WiFi transmission channel \{$1$, $6$, $11$\},  number of nodes simultaneously receiving data from the transmitter $N$ (from $1$ to $4$), and distance between the transmitter and the receivers \{$1$ m, $5$ m, $10$ m\}. 
For each configuration, we measured the transmission duration of a $100$ MB file (the average size of a music video in full HD resolution). Each configuration of the network has been tested $10$ times, and the experimental campaign lasted more than $45$ days, for a total of $M = 5400$ experiments. The network is deployed following a single-hop star topology, with a transmitting central node, called Access Point, and one or more nodes waiting for data reception.

The dataset has been divided into a training set $(\mathbf{X}_{tr}, \mathbf{y}_{tr})$ and a test set $(\mathbf{X}_{test}, \mathbf{y}_{test})$: $80\%$ of the data was assigned to the former, leaving the remaining $20\%$ for performance testing, in order to take into account the trade-off between learning quality and classification accuracy. In particular, the classification performance was tested using four subsets of the network parameters as input features, in order to measure how a higher number $n$ of known parameters could improve the final classification performance: the first subset contains only the ETA, whereas the others include either the ETA and the transmission power ($P_{Tx}$) or the ETA and the distance $d$, or all three features (ETA, $P_{Tx}$, $d$). All these network parameters are available at the receiving nodes, thus no modification of the standard signaling protocol is needed. Each value in either $\mathbf{y}_{tr}$ or $\mathbf{y}_{test}$ is the number $N$ of nodes actively receiving data from the AP, and is associated to its corresponding input feature vector in $\mathbf{X}_{tr}$ or $\mathbf{X}_{test}$.

Finally, we tried to simulate a real case scenario, where we assess the classification performance in a more crowded environment, \emph{e.g.}, a domestic location, in which the distribution of simultaneously receiving nodes is shifted towards $N = 4$ nodes. We introduced this case to study how the different classification algorithms are affected by this choice and whether a change in the parameters of the classification model leads to a better performance.

\section{Overview of ML techniques}\label{sec:ml}
In this section, we first describe the structure of the dataset we used to collect data from the experiments and later introduce the classification techniques adopted to predict the number of users simultaneously receiving data from the AP.

\subsection{Data structure}\label{subsec:dataStructure}
%(containing, for instance, the ETA value associated with a transmission) 
The dataset adopted in this work is composed by a matrix of input data $\mathbf{X}\in \mathbb{R}^{M \times n}$ and a vector of outputs $\mathbf{y} \in \mathbb{R}^M$. The $i$-th row in $\mathbf{X}$ and the $i$-th element in $\mathbf{y}$ define a pair $(\mathbf{x}^{(i)}, y^{(i)})$, for $i=1,\dots,M$, where $\mathbf{x}^{(i)} \in \mathbb{R}^n$ is the input feature vector and $y^{(i)}$ is the label representing the value of $N$ related to $\mathbf{x}^{(i)}$ (\emph{i.e.}, the number of active nodes during that transmission). We used different supervised classification techniques to identify the model that best associates a given subset of the examples in $\mathbf{X}$ to a set of labels, pairing each example $\mathbf{x}$ to a class. %To avoid overfitting, which occurs when a model starts to memorize data rather than learn to generalize from trends,
We split the dataset into a training set $(\mathbf{X}_{tr} \in \mathbb{R}^{m \times n}, \mathbf{y}_{tr} \in \mathbb{R}^m)$ used to train and optimize the learning algorithms, and a test set $(\mathbf{X}_{test} \in \mathbb{R}^{(M-m) \times n}, \mathbf{y}_{test} \in \mathbb{R}^{M-m}$, to evaluate the performance of each classifier. 

\subsection{Naive Bayes classifier}\label{subsec:Bayesian}

The Bayesian classifier~\cite{Duda} is a probability model that computes the posterior probability $p(y_k|\mathbf{x})$ of a class given an input example $\mathbf{x}$, for each of the $k$ possible classes. Using Bayes' rule, this probability is computed as
\begin{align}
p(y_k|\mathbf{x}^{(i)}) \propto p(\mathbf{x}^{(i)}|y_k)p(y_k) \ , \ i=1,\dots,m,
\end{align}
where the posterior probability is proportional to the likelihood $p(\mathbf{x}^{(i)}|y_k)$ and the prior $p(y_k)$. The object $\mathbf{x}^{(i)}$ is classified under the class that holds the highest posterior probability: 
\begin{align}\label{eq:MAP}
y^*_{\text{MAP}} = \argmax_k p(\mathbf{x}^{(i)}|y_k)p(y_k),
\end{align}
known as Maximum A Posteriori (MAP) formulation. 
The likelihood expresses how probable the input data $\mathbf{x}^{(i)}$ is for a given class $y_k$, whereas the prior captures the assumptions about the class, before observing the data, in the form of a prior probability distribution $p(y_k)$. If no prior knowledge about the class distribution is available, a uniform prior $p(y_k)=c$ is assumed and \eqref{eq:MAP} turns into the Maximum Likelihood (ML) solution $y^*_{\text{ML}} = \argmax_k p(\mathbf{x}^{(i)}|y_k)$.
The likelihood function can be modeled using different probability distribution functions (\textit{pdf}s): in this paper we used both a Gaussian and a Poisson \textit{pdf}, setting their parameters (mean and variance for the former, mean for the latter) according to the examples in the training set. In particular, we ran the classification algorithm with input examples $\mathbf{x}^{(i)}$ of different dimensions, based on the number of features $n$ involved: the idea behind this approach is to analyze how additional knowledge on an experiment  impacts on the performance of the classifier. Therefore, the input vector contains either the ETA alone, or the pair (ETA, $P_{Tx}$) or (ETA, distance), or the triple (ETA, $P_{Tx}$, distance), and different \textit{pdf}s are generated accordingly. For instance, for the second case ($n=2$) with ETA and distance ($d$) as input features, \eqref{eq:MAP} is reformulated as 
\begin{align}
y^*_{\text{MAP}} = \argmax_k p(ETA^{(i)}|y_k,d^{(i)})p(d^{(i)})p(y_k), \ 
\end{align}
$
i=1,\dots,m,
$
where $p(d^{(i)}|y_k)=p(d^{(i)})$ for a homogeneous dataset.

\subsection{Support Vector Machines}\label{subsec:SVC}

Support Vector Machines (SVMs)~\cite{Cortes1995} identify the hyperplane better separating the classes in a multi-dimensional feature space. A generic hyperplane can be written as the set of points $\mathbf{x}$ satisfying $\mathbf{w}^T \mathbf{x} - b = 0$, where $\mathbf{w}$ is the normal vector to the hyperplane and $b$ represents the offset from the origin along $\mathbf{w}$. When the data points are linearly separable, it is possible to identify a pair of hyperplanes able to perfectly separate all the objects. By defining these hyperplanes as $\mathbf{w}^T \mathbf{x} - b = \pm 1$, the margin, \emph{i.e.}, the region in-between, has length $\frac{2}{||\mathbf{w}||}$, and the objective is to maximize $\frac{1}{||\mathbf{w}||}$. If we consider a binary classification problem, $y^{(i)} \in \{-1,1\}$, the maximization of $\frac{1}{||\mathbf{w}||}$ is equivalent to the minimization of $\frac{1}{2}||\mathbf{w}||^2$; therefore the problem can be formulated as:
\begin{align}\label{eq:SVM_optprob1}
&\min_{\mathbf{w},\mathbf{b}} && \frac{1}{2} ||\mathbf{w}||^2 \nonumber\\
& \text{s.t.} &&\  y^{(i)}(\mathbf{w}^T\mathbf{x}^{(i)} + b) \ge 1,\ i=1,\dots,m,
\end{align}
where the constraint $y^{(i)}(\mathbf{w}^T\mathbf{x}^{(i)} + b) \ge 1$ has been added to guarantee that all $\mathbf{x}^{(i)}$s lie outside the margin.
Equation \eqref{eq:SVM_optprob1} assumes that all training examples can be separated, which is not verified in general. As a result, \eqref{eq:SVM_optprob1} can be modified to allow for some tolerance in the classification error. Therefore, for each training example $\mathbf{x}^{(i)}$, there is an associated non-negative slack-variable $\xi_i \ge 0$, $i=1,\dots,m$, allowing each example to lie on the other side of the hyperplane separating its class. The optimization problem then becomes:
\begin{align}\label{eq:SVM_optprob2}
& \min_{\mathbf{w},\mathbf{b}} && \frac{1}{2} ||\mathbf{w}||^2 + C \sum_{i=1}^{m} \xi_i \nonumber\\
& \text{s.t.} &&  y^{(i)}(\mathbf{w}^T\mathbf{x}^{(i)} + b) \ge 1 - \xi_i,\ \xi_i \ge 0,\ i=1,\dots,m,
\end{align}
where the parameter $C$ balances the trade-off between having a large margin and ensuring that most examples lie in the region associated to their class. The ``dual'' formulation of Eq.~\eqref{eq:SVM_optprob2} is
\begin{align}\label{eq:SVM_optprob3}
&\max_{\pmb{\alpha}} & &\sum_{i=1}^{m} \alpha_i - \frac{1}{2} \sum_{i,j=1}^{m}y^{(i)}y^{(j)}\alpha_i \alpha_j (\mathbf{x}^{(i)})^T \mathbf{x}^{(j)} \nonumber\\
&\text{s.t.} & & 0 \le \alpha_i \le C,\ \ i=1,\dots,m, \nonumber\\
& && \sum_{i=1}^{m}\alpha_i y^{(i)}=0, 
\end{align}
where $\pmb{\alpha}=(\alpha_1,\dots,\alpha_m)$ is the vector of Lagrange multipliers. It can be shown that 
\begin{align}\label{eq:SVM_weights}
\mathbf{w} = \sum_{i=1}^m \alpha_i y^{(i)} \mathbf{x}^{(i)}, 
\end{align}
by which it is possible to calculate the optimal weights in terms of the optimal values of $\pmb{\alpha}$. Now, a prediction for a new example input $\mathbf{x}$ can be performed calculating $\mathbf{w}^T \mathbf{x}+b$, and predicting $y=1$ (or $y=-1$) if this quantity is bigger (or smaller) than zero. Hence, the prediction function can be written as 
\begin{align}\label{eq:SVM_predfunlin}
f_\mathbf{w}(\mathbf{x}) = sgn \left( \sum_{i=1}^m \alpha_i y^{(i)} (\mathbf{x}^{(i)})^T\mathbf{x} +b \right),
\end{align}
which only depends on the inner product between the input vector $\mathbf{x}$ and the subset of training vectors $\mathbf{x}^{(i)}$ for which $\alpha_i \ne 0$. Moreover, from \eqref{eq:SVM_optprob3} it turns out that this subset only contains those training examples, known as Support Vectors (SVs), lying within the margin or in the region of the hyperspace belonging to the other class: as a consequence, the actual complexity of the prediction process only depends on the number of support vectors. 
%At the beginning of this section we assumed that the training examples were linearly separable: actually, this constraint can be easily relaxed, providing better generalization by identifying more accurate hyperplanes able to non-linearly separate the data. This can be done by noting that in \eqref{eq:SVM_predfunlin} SVs only appear inside scalar products \cite{Boser1992}, and \eqref{eq:SVM_weights} need not be calculated explicitly. Therefore, 
The inner product $(\mathbf{x}^{(i)})^T \mathbf{x} + b$ in \eqref{eq:SVM_predfunlin} can be replaced by particular non linear functions $k(\mathbf{x}^{(i)},\mathbf{x})$, known as \textit{kernels}, which correspond to scalar products between either linear or non linear transformations of $\mathbf{x}^{(i)}$ and $\mathbf{x}$. Substituting $k(\mathbf{x}^{(i)},\mathbf{x}^{(j)})$ in \eqref{eq:SVM_optprob2} and $k(\mathbf{x}^{(i)},\mathbf{x})$ in \eqref{eq:SVM_predfunlin}, we thus obtain the optimal prediction function in a non-linear feature space, rather than in input space:
\begin{align}
f_\mathbf{w}(\mathbf{x}) = sgn \left( \sum_{i=1}^{m} \alpha_i y^{(i)} k(\mathbf{x}^{(i)},\mathbf{x}) \right).
\end{align}

Finally, as the classification problem studied in this paper involves $|C|=4$ classes, we adopted a generalization of the standard SVM known as multiclass SVM~\cite{Bishop2006}, which works by reducing the multiclass problem into a number of binary problems. 
This generalization, known as one-vs-one classification, works by building a set of $|C|(|C|-1)/2$ binary classifiers, and then selecting the class that is assigned by the majority of the classifiers.

\subsection{k-Nearest Neighbor}
In k-Nearest Neighbor (k-NN), an example $\mathbf{x}$ is classified under the most common class among its $k$ nearest neighbors: $k$-NN is a type of lazy learning, since a learning phase is not needed at all. 
The neighbors are taken from a set of examples for which the class is known: this set can be thought of as the training set for the algorithm.
In other words, the training phase of the algorithm simply consists in storing the training examples: once a test example is provided, the algorithm classifies it by assigning the most frequent class among its $k$ nearest training examples. The parameter $k$ is chosen so as to balance the trade-off between reducing the effect of noise (large $k$) and avoiding the creation of indistinct boundaries between the classes (low $k$). %When $k = 1$, the algorithm is generally known as Nearest Neighbor (NN).

\section{Experimental Results}\label{sec:results}

In this section, we present the classification results of the four algorithms: Naive Bayes (NB), linear Support Vector Machines (SVM-L), radial Support Vector Machines (SVM-R)\footnote{Radial kernels are one of the most effective kernel implementations.}, k-Nearest Neighbor (kNN). 
The results are split into two parts: in the first part, we test the four classifiers on the whole dataset; in the second part, we train the classifiers on two subsets of data which mimic a more realistic scenario with unbalanced classes. In both cases (balanced and unbalanced training and test), we divide the dataset into 5 folds and train/test the classifiers 5 times, \emph{i.e.}, using 4 folds for training and the remaining fold for testing, every time. For the second part, we maintain the same 5 folds and sample a subset of the training and test data in order to produce an unbalanced dataset with more examples of transmission with multiple nodes (higher value of $N$). Two training/test subsets have been created with different proportion of examples per class, see Table~\ref{tab:dataset} for the details. For instance, subset ``10-20-50-100" means that the first class ($N=1$) is subsampled taking only $10\%$ of the data, the second class ($N=2$) maintains $20\%$ of the data, \emph{etc}.

\begin{table}[th]
	\caption{Number of examples per class.}
	\label{tab:dataset}
	\begin{center}
		\begin{tabular}{c | r | r | r | r }
			Subset & N = 1 &  N = 2  &  N = 3  &   N = 4 \\
			\hline
			Full dataset	& 1,350 & 1,350 & 1,350 & 1,350 \\
%			\hline
			10-20-50-100	& 135 & 270 & 675 & 1,350 \\
%			\hline
			20-30-40-100 & 270 & 405 & 540 & 1,350 \\
		\end{tabular}
	\end{center}
		\vspace{-0.275cm}
\end{table}%

Since we deal with a multi-class classification problem, we adopt the $F_1$ score\footnote{$F_1$ score (also F-score or F-measure) is a common accuracy metric.} as a measure of performance for each class~\cite{DBLP:journals/ipm/SokolovaL09}:
\begin{equation}
F_1 = \frac{2 \cdot P \cdot R} {P + R}
\end{equation}
where $P$ is the precision, \emph{i.e.},  the class agreement of the data labels with the positive labels given by the classifier, and $R$ is the recall, that is the effectiveness of a classifier in identifying positive labels~\cite{han2011data}. In addition, we analyze the performance in terms of Receiver Operating Characteristic (ROC) curve which shows the fraction of true positive decisions of a classifier for a given rate of false positive decisions~\cite{Bradley19971145}.

\subsection{Full dataset}\label{subsec:fulldataset}
\begin{figure}[t]
	\centering
	\includegraphics[width=\columnwidth]{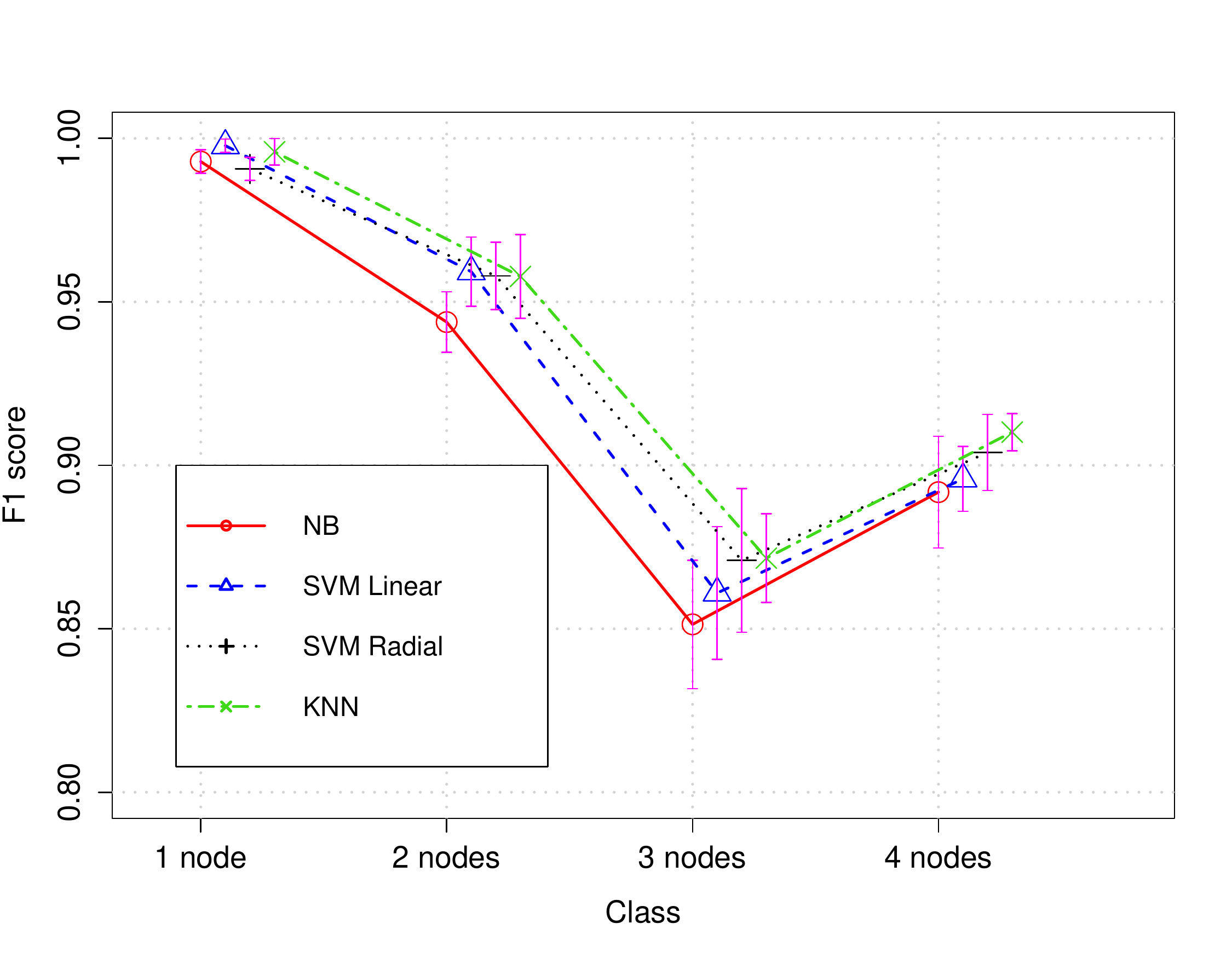}
	\caption{Performance (mean and standard deviation) on full dataset.}
	\label{fig:full}
		\vspace{-0.275cm}
\end{figure}

The classification performance for the full dataset is shown in Figure~\ref{fig:full}. For each class (number of nodes $N=\{1, \dots, 4\}$), we plot the average and the standard deviation of the $F_1$ score of each classifier when all the features are used (ETA, $P_{Tx}$, and distance). In general, the performance is very high and it is possible to identify a common trend for all the classifiers: the 3-node class is harder to classify, mainly because the ``tails'' of this class overlap with the 2-node and 4-node classes. Table~\ref{tab:full} summarizes the performance for each class and the total average $F_1$ score given a particular set of features. As we add more evidence (in terms of features), the performance of each classifier increases. 

\begin{figure}[t]
	\centering
	\includegraphics[width=\columnwidth]{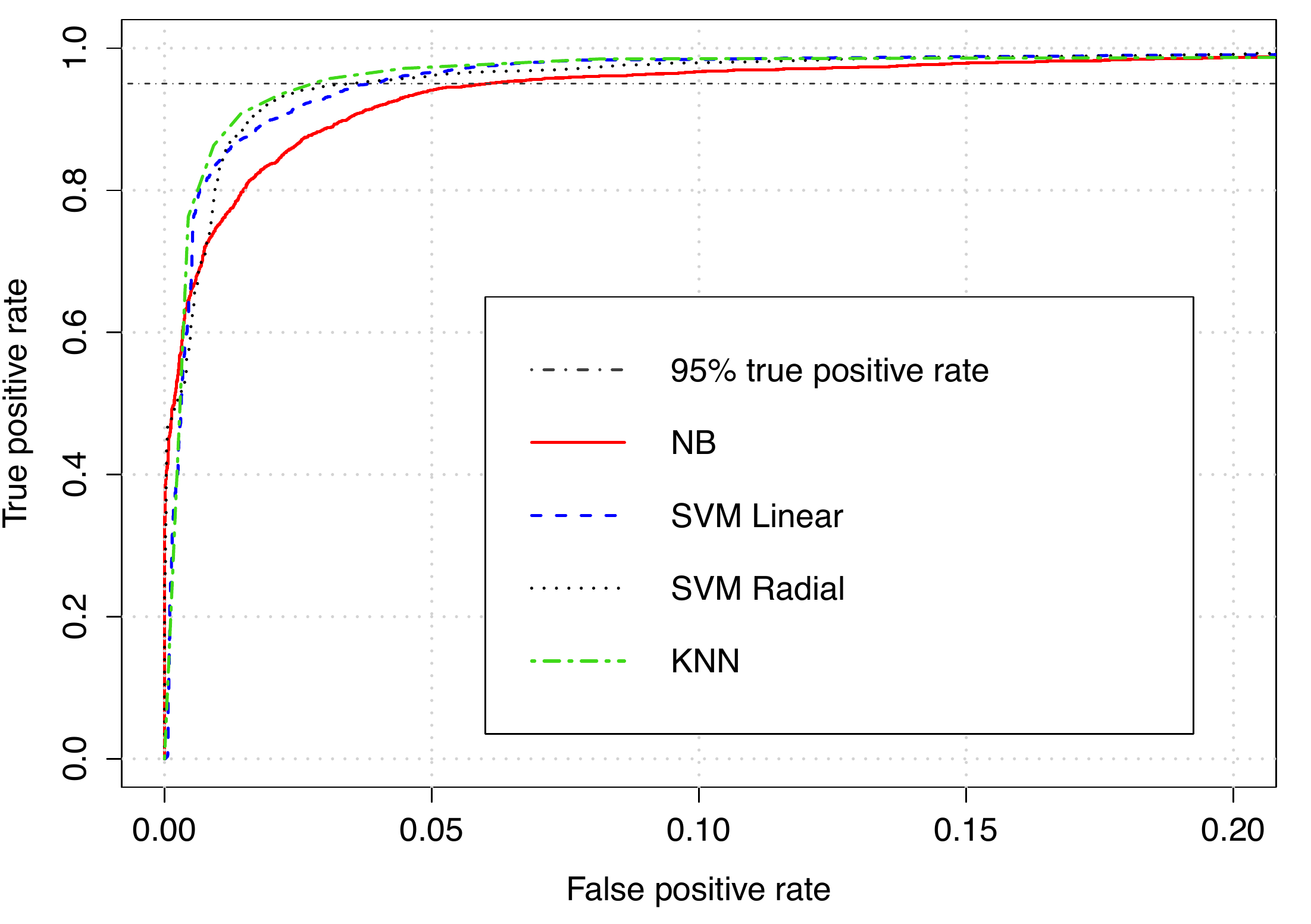}
	\caption{ROC curves on full dataset.}\label{fig:ROC}
%	\vspace{-0.5cm}
\end{figure}

In \figurename~\ref{fig:ROC}, we show the analysis of the classification performance in terms of the ROC curve. This curve can be used to select the suitable operating point of the classifier; in particular, we see that at 20\% rate of false positives, all the classifiers achieve an almost perfect performance (dashed line indicates 100\% true positives). It is also worth noting that in the range 3\% - 6\% of false positive rate, all the classifiers are above the 95\% true positive rate (dotted line). Table~\ref{tab:ROC} shows the cutoff points for the false positive rate that is required to achieve at least 95\% of true positive rate. \figurename~\ref{fig:ROC} and Table~\ref{tab:ROC} confirm that the NB approach is the slowest to reach the best performance.

\begin{table}[th]
	\caption{$F_1$ score performance on full dataset.}
	\label{tab:full}
	\begin{center}
		\begin{tabular}{ c | l | c | c | c | c || c }
			Features & Classifier & $N = 1$ & $N = 2$ & $N = 3$ &  $N = 4$ & Average\\
			\hline
			\multirow{4}{*}{ETA} & NB & 0.986 & 0.929 & 0.818 & 0.863 & 0.899 \\
%			\cline{2-7}
			& SVM-L	& 0.996 & 0.952 & 0.845 & 0.884 & 0.919 \\
%			\cline{2-7}
			& SVM-R	& 0.997 & 0.950 & 0.841 & 0.888 & 0.919 \\
%			\cline{2-7}
			& kNN	& 0.997 & 0.948 & 0.825 & 0.875 & 0.911 \\
			\hline
			\hline
			\multirow{4}{*}{ETA, $d$} & NB    & 0.990 & 0.917 & 0.813 & 0.877 & 0.899 \\
%			\cline{2-7}
			& SVM-L & 0.997 & 0.953 & 0.840 & 0.877 & 0.917 \\
%			\cline{2-7}
			& SVM-R & 0.996 & 0.960 & 0.861 & 0.899 & 0.929 \\
%			\cline{2-7}
			& kNN & 0.997 & 0.960 & 0.854 & 0.892 & 0.926 \\
			\hline
			\hline
			\multirow{4}{*}{ETA, $P_{Tx}$} & NB     & 0.989 & 0.939 & 0.84 & 0.879 & 0.912 \\
%			\cline{2-7}
			& SVM-L & 0.997 & 0.958 & 0.851 & 0.887 & 0.923 \\
%			\cline{2-7}
			& SVM-R & 0.992 & 0.957 & 0.842 & 0.876 & 0.916 \\
%			\cline{2-7}
			& kNN & 0.996 & 0.958 & 0.842 & 0.879 & 0.919 \\
			\hline
			\hline
			\multirow{4}{*}{ETA, $P_{Tx}$, $d$} & NB     & 0.993 & 0.944 & 0.851 & 0.892 & 0.920 \\
%			\cline{2-7}
			& SVM-L & 0.998 & 0.959 & 0.861 & 0.896 & 0.928 \\
%			\cline{2-7}
			& SVM-R & 0.991 & 0.958 & 0.871 & 0.904 & 0.931 \\
%			\cline{2-7}
			& kNN & 0.996 & 0.958 & 0.872 & 0.910 & 0.934 \\
		\end{tabular}
				\vspace{-0.275cm}
	\end{center}
\end{table}%

\begin{figure}[t]	
	\centering
	\includegraphics[width=\columnwidth]{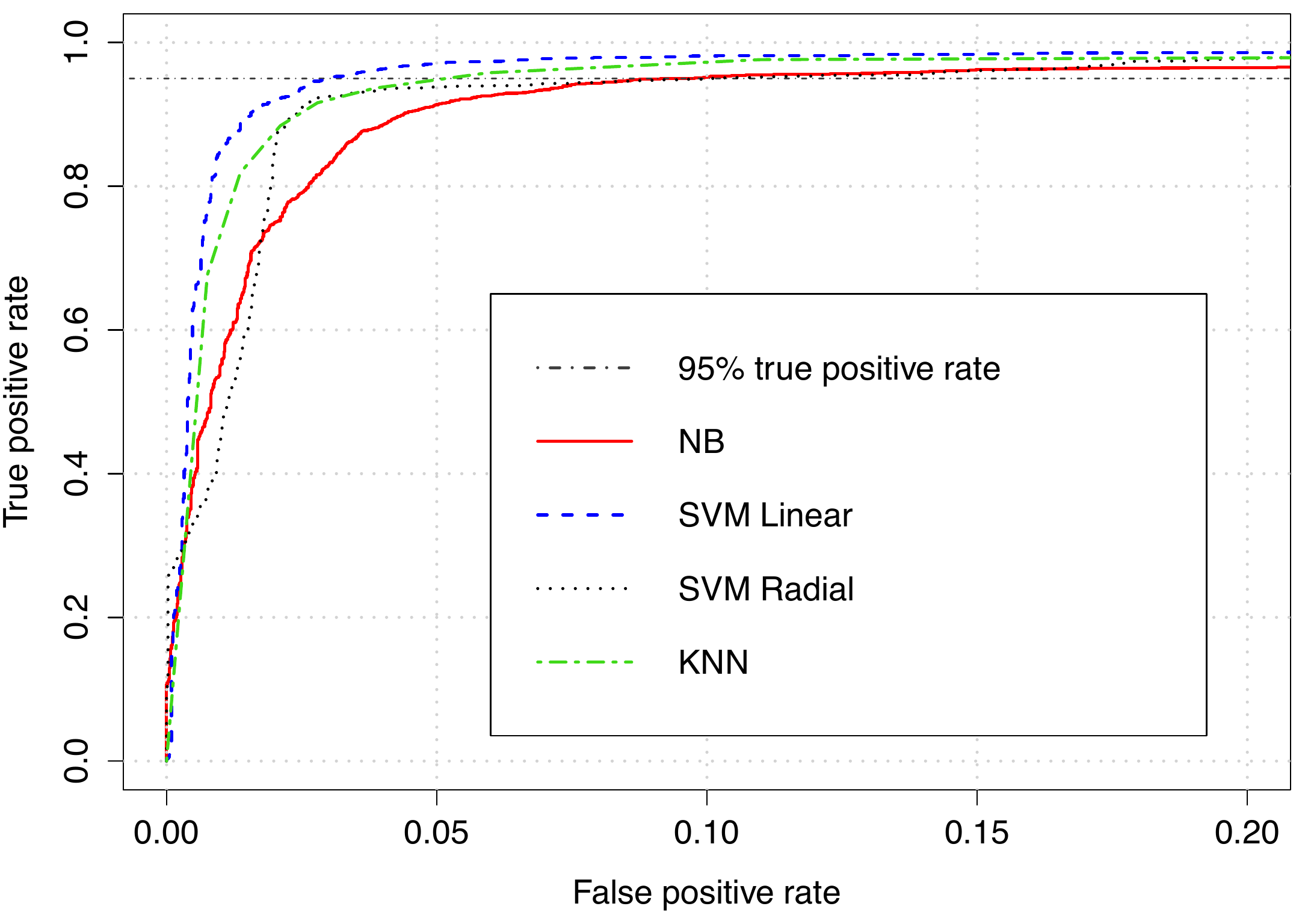}
	\caption{ROC curves on subset 20-30-50-100.}\label{fig:ROC20}
%	\vspace{-0.5cm}
\end{figure}

\begin{figure}[t]
	\centering
	\includegraphics[width=\columnwidth]{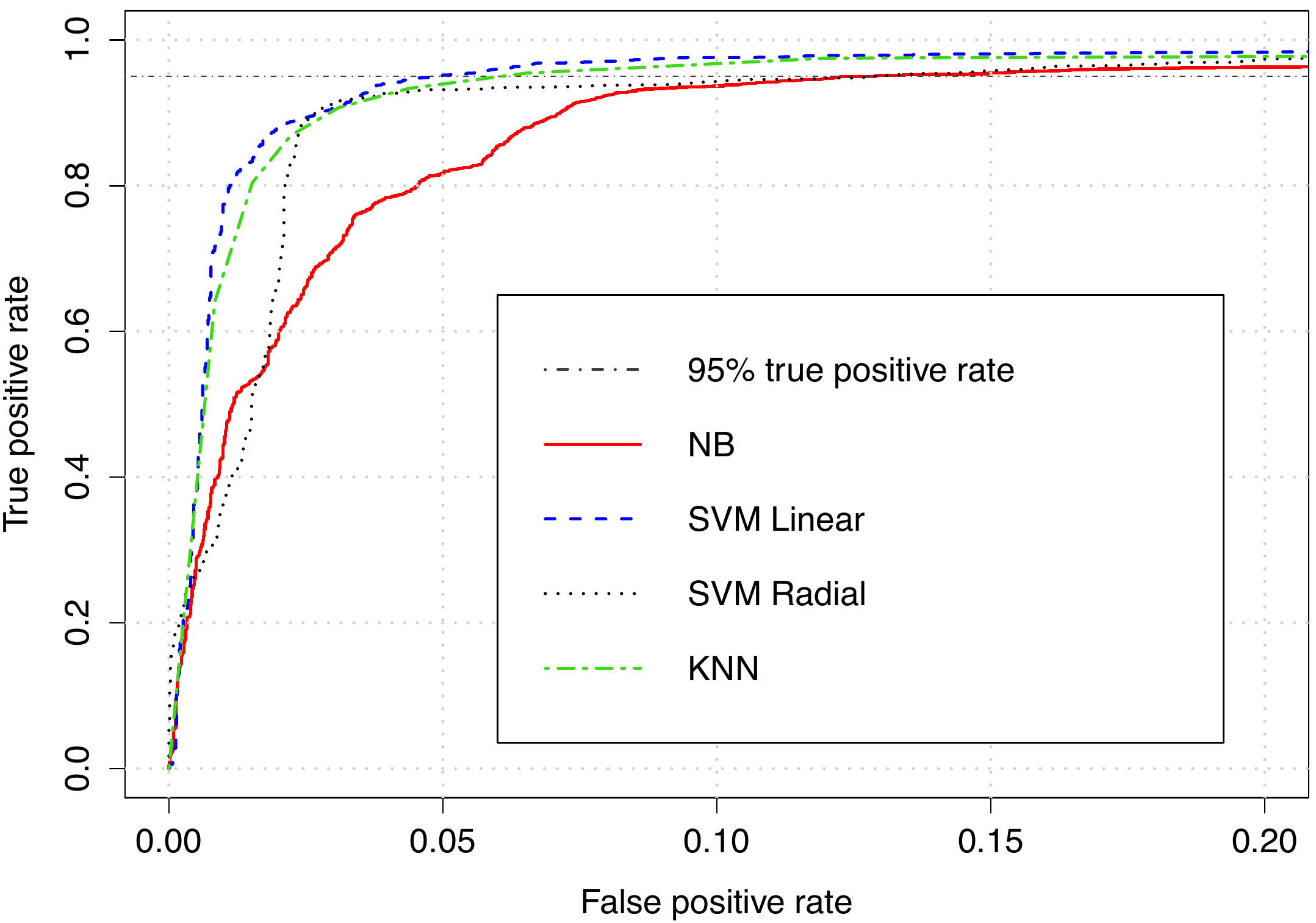}
	\caption{ROC curves on subset 10-20-30-100.}\label{fig:ROC10}
%	\vspace{-0.5cm}
\end{figure}

\begin{table}[th]
	\caption{False positive rate @ true positive rate = 0.95.}
	\label{tab:ROC}
	\begin{center}
		\begin{tabular}{ l | c | c | c }
			& ETA, $P_{Tx}$, $d$ & \multicolumn{2}{c}{ETA} \\
			Classifier & Full dataset &  10-20-30-100 & 20-30-50-100 \\
			\hline
			NB (prior) & 0.060 & 0.094 & 0.129 \\
%			\hline
			SVM-L (no weight) & 0.046 & 0.030 & 0.048  \\
%			\hline
			SVM-R (no weight) & 0.035 & 0.101 & 0.128  \\
%			\hline
			kNN & 0.030 & 0.060 & 0.065 \\			
		\end{tabular}
				\vspace{-0.275cm}
	\end{center}
\end{table}%

\subsection{Unbalanced dataset}
In this subsection, we study the classification performance when classes are unbalanced and only one feature (the ETA) is available. In particular, we present the results of two subsamples of the classes with the following proportions (compared to the full dataset): i) $10\%$, $20\%$, $30\%$, $100\%$, ii) $20\%$, $30\%$, $50\%$, $100\%$, respectively for the 1-node, 2-node, 3-node, and 4-node classes.
We tested the NB and SVM with two different approaches: one that incorporates the knowledge of the disproportion among classes, and another that does not take into account this information. For the NB, we trained NB with a uniform (NB uniform) distribution over the classes, so as to simulate a lack of knowledge about the unbalanced class distribution. We also trained a NB which computes the prior distribution $p(y_k)$ over the classes based on the relative frequency of class $y_k$ in the training set (NB prior). 
For the SVM, both the linear (SVM-L) and radial kernel (SVM-R) were used, training the algorithm either without modifying the weights of the classes (SVM no weight), or setting the values of the weights proportionally to the amount of training instances of each class in the training set (SVM weight), so that the $C$ parameter in \eqref{eq:SVM_optprob2} and \eqref{eq:SVM_optprob3} takes into consideration the unbalanced data by setting the same sum of weights for each class. Finally, the k-NN algorithm was executed without any modification.

\begin{table}[th]
	\caption{$F_1$ score performance on subset 10-20-30-100.}
	\label{tab:10}
	\begin{center}
		\begin{tabular}{ l | c | c | c | c || c }
			Classifier & $N = 1$ & $N = 2$ & $N = 3$ &  $N = 4$ & Average\\
			\hline
			NB (uniform) & 0.985 & 0.907 & 0.781 & 0.893 & 0.891\\
%			\hline
			NB (prior) & 0.985 & 0.942 & 0.830 & 0.925 & 0.920  \\
%			\hline
			SVM-L (no weight) & 1.000 & 0.953 & 0.837 & 0.929 & 0.930 \\
%			\hline
			SVM-L (weight)& 1.000 & 0.948 & 0.830 & 0.918 & 0.924 \\
%			\hline
			SVM-R (no weight) & 0.989 & 0.941 & 0.836 & 0.929 & 0.924 \\
%			\hline
			SVM-R (weight)& 0.989 & 0.935 & 0.830 & 0.919 & 0.918 \\
%			\hline
			kNN & 1.000 & 0.955 & 0.827 & 0.926 & 0.927 \\
			
		\end{tabular}
	\end{center}
\end{table}%

\begin{table}[th]
	\caption{$F_1$ score performance on subset 20-30-50-100.}
	\label{tab:20}
	\begin{center}
		\begin{tabular}{ l | c | c | c | c || c }
			Classifier & $N = 1$ & $N = 2$ & $N = 3$ &  $N = 4$ & Average\\
			\hline
			NB (uniform) & 0.991 & 0.924 & 0.737 & 0.895 & 0.887 \\
%			\hline
			NB (prior) & 0.989 & 0.928 & 0.775 & 0.938 & 0.907 \\
%			\hline
			SVM-L (no weight) & 1.000 & 0.953 & 0.837 & 0.929 & 0.930 \\
%			\hline
			SVM-L (weight) & 1.000 & 0.948 & 0.83 & 0.918 & 0.924 \\
%			\hline
			SVM-R (no weight) & 0.993 & 0.947 & 0.800 & 0.938 & 0.919 \\
%			\hline
			SVM-R (weight)& 0.994 & 0.939 & 0.798 & 0.929 & 0.915 \\
%			\hline
			kNN & 0.998 & 0.95 & 0.795 & 0.935 & 0.920 \\
			
		\end{tabular}
	\end{center}
\end{table}%

In Table~\ref{tab:ROC}, we added the false positive rate cutoffs for each classifier. Since these two experiments, due to a reduced amount of training examples, constitute harder classification tasks with respect to the balanced-class setting, the values of the cutoff points are, in general, higher. This means that, in order to achieve a 95\% true positive rate, it is at least needed to double the false positive error rate. It seems that SVM-L is much less prone to this type of performance decay. We will further investigate this particular behavior of the SVM with linear kernel.
The results in Table~\ref{tab:10} and Table~\ref{tab:20} show that, in general,
%the ML techniques taking into account the disequilibrium among the classes in the unbalanced dataset enjoy slight performance improvements. In particular, 
SVM-L, achieving the highest $F_1$-score values in most classification tasks, seems to be preferable for the unbalanced datasets, whereas NB provides inferior results. In addition, kNN could be a good choice overall, as it attains good results at the lowest computational cost (as no training phase is needed).

\subsection{Discussion and further analyses}

The results of the experiments presented in this paper are very promising. The performance of the classifiers in all the considered configurations are high on average ($F_1$ score above 0.89) if we consider that only three features were used (namely ETA, $P_{Tx}$, and $d$); moreover, the study of the ROC curve shows that a true positive rate equal to 95\% can be achieved with a very small false positive rate (around 2-5\%). We also observed a slight increment in performance when going from a single feature (ETA) to three features (ETA, $P_{Tx}$ and $d$), but this increment in performance is not statistically significant. Instead, a substantial difference between the NB classifier and the SVM and k-NN classifiers has been observed\footnote{We performed a paired t-test significance analysis with $\alpha = 0.05$.}. In the second part of the experimental analysis, we showed that the classification performance continues to be high even when classes are not balanced and the information about the skewness is not taken into account. Unexpectedly, weighted SVM performed worse than the unweighted (default) version. We are going to investigate further this problem in the future.

%\section{New results}
\begin{table}[t] 
	\caption{Percentage mean prediction errors.}\label{tab:ETA_pred}
	\tabcolsep=0.11cm 
	\begin{tabular}{c|c|c|c|c} 
%		\hline 
		& $N_{pred} = 1$ & $N_{pred} = 2$ & $N_{pred} = 3$ & $N_{pred} = 4$ \\  
		\hline 
		$N_{real} = 1$ & $9.57 \pm 4.68$ & $89.90 \pm 16.55$ & $179.19 \pm 26.68$ & $282.57 \pm 42.20$ \\ 
		$N_{real} = 2$ & $46.76 \pm 5.79$ & $6.38 \pm 5.26$ & $47.90 \pm 16.19$ & $103.92 \pm 29.58$ \\ 
		$N_{real} = 3$ & $63.96 \pm 4.63$ & $32.63 \pm 7.86$ & $6.22 \pm 4.42$ & $38.23 \pm 12.83$ \\ 
		$N_{real} = 4$ & $73.42 \pm 2.56$ & $50.54 \pm 5.09$ & $26.55 \pm 7.27$ & $3.85 \pm 2.82$ \\ 
%		\hline 
	\end{tabular} 
\end{table} 

\begin{table}[!t] 
	\caption{Distribution of $\mathbb{P}[N_{pred}|N_{real}]$ on full dataset, with ETA as input feature and SVR-R as ML technique.}\label{tab:prob_distr}
	 	\begin{tabular}{c|c|c|c|c} 
		%		\hline 
		& $N_{pred} = 1$ & $N_{pred} = 2$ & $N_{pred} = 3$ & $N_{pred} = 4$ \\  
		\hline 
		$N_{real} = 1$ & 0.9937 & 0.0044 & 0.0012 & 0.0006 \\ 
		$N_{real} = 2$ & 0.0035 & 0.9138 & 0.0718 & 0.0109 \\ 
		$N_{real} = 3$ & 0.0033 & 0.0679 & 0.7663 & 0.1625 \\ 
		$N_{real} = 4$ & 0.0037 & 0.0186 & 0.1864 & 0.7912 \\ 
		%		\hline 
	\end{tabular} 
\end{table} 

We claim that the exact number of users in the network is an important piece of information, as the predicted ETA is highly dependent on the value of $N$. To prove this, in Table~\ref{tab:ETA_pred} we show percentage mean prediction errors on the ETA, as a function of the number of users given as input, computed using the best-performing ML technique in \cite{deltesta_globecomm}. %Each experiment in the test set consists of an ETA measured value and a number of features to be used for the ETA prediction, including the number of users.
In the table, we indicate with $N_{real}$ the actual number of active nodes during a given experiment, and with $N_{pred}$ the value of $N$ given as input to the ETA predicted by the algorithm. The goal is to compare the ETA prediction errors when an erroneous number of nodes is used instead of the right one. For each experiment in the test set, the ETA has been predicted with a given value of $N_{pred}$, and a percentage error has been computed with respect to the measured ETA value. The average of all the errors and standard deviations for a given $(N_{real}, N_{pred})$ pair is reported in Table~\ref{tab:ETA_pred}. Note that the errors on the diagonal are related to the predictions performed using the ``right'' value of $N$ ($N_{pred} = N_{real}$); as expected, these values are the lowest for each class of experiments. The worst case is when the algorithm predicts a number of nodes greater than 1, $N_{pred} > 1$, when $N_{real} = 1$. In addition, when $N_{real}=2$ (or $N_{real}=3$), using the nearest value of $N$ for the prediction, \emph{i.e.}, either $N_{pred}=1$ or $N_{pred}=3$ (or $N_{pred}=2$ or $N_{pred}=4$, respectively), gives approximately symmetric mean percentage errors. By computing the distribution $\mathbb{P}[N_{pred}=k|N_{real}=n]$ with $k,n \in [1, 2, 3, 4]$, see Table~\ref{tab:prob_distr}, it is possible to obtain the percentage weighted average prediction errors $\delta_i, \ i \in [1, 2, 3, 4]$, for each value of users $N_{real}$. For the aforementioned case, this computation gives $\bm{\delta} = \{10.3, 10.5, 13.4, 9.2\}$ which means that, on average, the percentage error is generally of the order of $10\%$, except when $N_{real}=3$, due to the overlap of the distribution of this class on the 2-node and 4-node ones, as already noted in Subsection~\ref{subsec:fulldataset}.

\section{Conclusions}
\label{sec:conclusions}
In this paper, we studied how the number of active nodes $N$, an important SDN parameter, can effectively be inferred using only the data available at each node of a wireless network. We studied the distribution of the error on the ETA of a WiFi transmission given a wrong prediction of this parameter, and how the use of ML techniques is important to analyze the information derived from the first transmitted file chunk and return the value of $N$. Based on this work, a further step could be to predict the number of active nodes on-the-fly, as a function of the amount of data already received. This could allow to simultaneously update both the ETA estimation and the number of nodes during a transmission.

%\section*{Acknowledgments}

\bibliographystyle{IEEEtran}
\bibliography{biblio}

\end{document}